\documentstyle[12pt,aasms4]{article}

\received{28 April 1997}

\lefthead{Combes \& Wiklind}
\righthead{Water at z$=$0.685}

\begin{document}

\title{Detection of water at z$=$0.685 towards B0218+357}
\author{Fran\c{c}oise Combes}
\affil{DEMIRM, Observatoire de Paris, 61 Av. de l'Observatoire,
F--75014 Paris, France; (bottaro@obspm.fr)}

\and

\author{Tommy Wiklind}
\affil{Onsala Space Observatory, S--43992 Onsala, Sweden; 
(tommy@oso.chalmers.se)}

\begin{abstract}

We report the detection of the H$_2$O molecule in absorption
at a redshift z$=$0.68466 in front of the gravitationally lensed
quasar B0218+357. We detect the fundamental transition of
ortho--water at 556.93\,GHz (redshifted to 330.59\,GHz).
The line is highly optically thick and relatively wide (15\,km/s FWHM),
with a profile that is similar to that of
the previously detected CO(2--1) and HCO$^+$(2--1) optically
thick absorption lines toward this quasar. From 
the measured level of the continuum at 330.59\,GHz, which
corresponds to  the level expected from the power--law spectrum
$S(\nu) \propto \nu^{-0.25}$ already observed at lower frequencies,
we deduce that the filling factor of the H$_2$O absorption is large.
It was already known from  the high optical thickness of the CO, $^{13}$CO and
C$^{18}$O lines that the molecular clouds entirely cover one of the
two lensed images of the quasar (all its continuum is absorbed);
our present results indicate that the H$_2$O clouds are covering a comparable
surface. The H$_2$O molecules are therefore not confined to small cores with a
tiny filling factor, but are extended over parsec scales.
The H$_2$O line has a very large optical depth, and only isotopic lines
could give us the water abundance.
We have also searched for the 183\,GHz line in absorption,
obtaining only an upper limit; this yields constraints on the
excitation temperature.
\end{abstract}


\section{Introduction}

Water is believed to be one of the most abundant molecules in the
interstellar medium (ISM). It can be formed through gas--phase chemistry
in cold, dense and thick clouds, with an abundance ratio  H$_2$O/H$_2$
between 10$^{-7}$ and 10$^{-5}$, depending on the chemical models,
the reactions rates used, the C/O abundance in the gas phase (Leung, Herbst 
\& Huebner 1984; Langer \& Graedel 1989);
in non--dissociative shocks the abundance ratio is calculated to be as 
high as 10$^{-4}$ (Draine, Roberge \&
Dalgarno 1983; Kaufman \& Neufeld 1996). The H$_2$O
abundance can also be enhanced through evaporation of grain mantles
in star--forming hot cores (Jacq et al. 1988; Brown, Charnley \& Millar 1988;
Gensheimer, Mauersberger \& Wilson 1996), so H$_2$O could play a major
role in the cooling of molecular clouds, and in the oxygen budget of the ISM.
Unfortunately, the broad atmospheric water lines prevent a direct
detection from the ground in our own Galaxy; up to now, 
no thermal emission from the main isotopomer in its fundamental lines
has been detected, and the H$_2$O abundance in the ISM remains poorly known.
Attempts have been made to determine the H$_2$O abundance
through observations of the isotopomers
HDO and H$_2^{18}$O (Henkel et al. 1987; Jacq et al. 1988, 1990; Wannier et al.
1991; Gensheimer et al. 1996), and through observations of the precursor
ion H$_3$O$^+$  (Phillips, van Dishoeck \& Keene 1992). Abundances of the
normal isotopomer of H$_2$O around  10$^{-5}$ have been deduced.
This is also confirmed by the detection in Orion of absorption lines at
2.66 microns with the Kuiper Airborne Observatory (KAO) (Knacke \& Larson
1991). The latter authors found an ortho--para ratio of 1, which
confirms that water had no time, after sublimation from grains,
 to reach the equilibrium high temperature ratio of 3.
Also the deuterated substitute HDO reveals a high degree of fractionation
(HDO/H$_2$O $=$ 100 D/H), which implies H$_2$O formation at low temperatures.

Since some of the H$_2$O submillimeter and FIR transitions are population
inverted, causing maser emission with a very high flux, they can be detected
from the ground even at 183\,GHz (Cernicharo et al. 1994).
Another method to avoid atmospheric absorption lines is to observe 
a remote object, for which the lines are redshifted outside the
broad atmospheric counterpart. Only one tentative detection has
been reported so far: the 752\,GHz para--water line in the
$z=2.28$ galaxy IRAS 10214+47 (Encrenaz et al. 1993; Casoli et al. 1994).
The emission in this remote starburst galaxy is apparently 
significantly enhanced by gravitational lensing.
Very recently, observations with the ISO satellite of the $2_{12}$ -- $1_{10}$
179.5 $\mu$m line of ortho--water in absorption against the continuum of 
the galactic center (SgrB2, Cernicharo et al. 1997) have revealed that the
H$_2$O molecule is abundant over very extended regions. 
It has also been detected in absorption in front of massive young stars
with strong IR continuum, around 6$\mu$m within the bending
vibration series (Helmich et al. 1996; van Dishoeck \& Helmich 1996).
Abundances of a few 10$^{-5}$ are deduced, with a tendency of scaling
with the amount of warm gas. Even higher abundances ($(3-8) \times 10^{-4}$)
have been derived from H$_2$O emission from the stellar wind of W Hya
with ISO--SWS and LWS, but the exact figures depend on the outflow
modelisation (Barlow et al. 1996; Neufeld et al. 1996). 

Here we report the first detection in absorption at high redshift of an
H$_2$O line. The high redshift allows us to avoid the high opacity
of the terrestrial atmosphere near the rest frequency, 
and because absorption is against a small continuum source, excellent spatial
resolution is achieved, equal to the angular size of the B0218+357 quasar core, 
which is only of the order of 1 milli-arcsec (Patnaik, Porcas \& Browne 1995).
At the distance of the absorber ($z=0.68466$, giving an angular
distance of 1089 Mpc, for $H_0$=75 km s$^{-1}$ Mpc$^{-1}$ and $q_0$ =0.5), 
this corresponds to only 5\,pc.

\section {Observations}

The observations were made with the IRAM 30m telescope at Pico Veleta near
Granada in Spain in March 1997. Table 1 displays the observational parameters. 
Several SIS receivers were used simultaneously: at 3mm, the receiver was tuned
to the H$_2$O 3$_{13}$-2$_{20}$ para line at 183\,GHz, redhifted to 108.811\,GHz
to obtain an estimate of the excitation temperature
(the lower level of this transition corresponds to a temperature of 
190\,K). The single--sideband (SSB) system temperature was 180\,K and the
rejection level of the image sideband $\sim$30\,dB. At 0.8mm, the H$_2$O 
1$_{10}$-1$_{01}$ ortho line at 557\,GHz, redshifted to 330.592 GHz, was 
observed. The receiver was operating in single--sideband, with a rejection 
level of a factor 4 (6\,dB measured on Orion);
its SSB receiver temperature was 90\,K and the system temperature was
between 400 and 2000\,K, depending on the atmospheric humidity, with 
an average of 700\,K. Two 512$\times$1\,MHz filterbanks and an
autocorelator backend were used. Here only the 1\,MHz resolution spectra
are presented, binned to a velocity resolution of a few km/s.

The observations were done with a nutating subreflector, with a beam--throw
of 1' in azimuth and a switching frequency of 0.5\,Hz.
The temperature scale was calibrated every 10 minutes by a chopper wheel on an
ambient temperature load, and on liquid nitrogen.
Pointing was checked on broadband continuum sources. The relative pointing
offsets between the 2  receivers
were of the order of 4". The frequency tunings and rejection levels were
checked by observing known molecular lines towards Orion, DR21 and IRC+10216.
The integration time was 8 and 20 hours on the 183\,GHz and 557\,GHz lines 
respectively, and a noise--level of 1.1 and 1.3\,mK in the T$_A^*$ antenna 
temperature scale was obtained, with a velocity resolution of 9\,km s$^{-1}$. 
The forward and beam efficiencies at each frequency are displayed in Table 1.

\smallskip

In order to derive the continuum flux, B0218+357 was observed regularly
with a continuum backend and in a fast switching mode (4 times higher
than in the line observing mode). The continuum level
from line observations obtained under good sky conditions was also used. 
The two estimates of the continuum flux agree.

The BL Lac object B0218+357 was selected for this first search for H$_2$O in
absorption because it is the absorbing system at high redshift with the
highest column density (Wiklind \& Combes 1995).
The remote quasar (z$_{\rm e}\approx$0.9, e.g. Browne et al. 1993) is
gravitationally lensed by a foreground  galaxy at z$=$0.68466, which
produces the absorption. The radio image of the quasar is composed of two
distinct flat--spectrum cores (A and B), with a small Einstein ring surrounding
the B image, of 335 milli--arcsecond (mas) in diameter (Patnaik, Browne \& King
1993). Since the ring has a steep spectrum, it is best interpreted as the 
image of a jet component, or a hot spot or knot in a jet that happens to lie
in the line of sight to the center of the lens. 
Owing to its steep spectral index, the Einstein ring gives a negligible
contribution to the continuum flux at millimeter wavelengths.
The intensity ratio between the two images is A/B $\approx$ 3--4 at several 
radio wavelengths, but the B--component has varied in flux by $\approx$ 
10\% in a few months (O'Dea et al. 1992; Patnaik et al. 1993).
Since the depth of the molecular absorption is less than the continuum level, 
but the absorption is optically thick, it follows that
the absorbing material does not cover the whole surface of the continuum
source. It is likely that only one image of the quasar is covered by
molecular clouds, since the two images A and B are separated by 1.8\,kpc
at the absorber distance. The fraction of the total continuum which 
is absorbed is $\approx$ 33\%.

\section{Results and discussion}

Figure \ref{h2o_f1} presents our H$_2$O detected spectrum, compared to
those of HCO$^+$(2--1) and CO(2--1) previously detected with the IRAM
30m telescope (Wiklind \& Combes 1995; Combes \& Wiklind 1995).
The linewidths are very similar, respectively 15, 16 and 15\,km/s 
for H$_2$O,  HCO$^+$(2--1) and CO(2--1), determined by gaussian fits.
This is a strong indication that the H$_2$O line is optically thick, as
the $^{13}$CO and C$^{18}$O isotopic lines have been detected,
with progressively reduced line--widths (Combes \& Wiklind 1995).
The 3$_{13}$-2$_{20}$ para line at 183\,GHz was not detected.
The 3$\sigma$ upper limit presented in Table\,1 were derived by assuming
the same linewidths for the two H$_2$O lines.

The redshift of the absorbing molecular gas, z$= 0.68466 \pm 0.00001$
(Wiklind \& Combes 1995), puts the redshifted H$_2$O(1$_{10}-1_{01}$)
line at 330.593\,GHz, which is close to the frequency of the
$^{13}$CO(3--2) transition at z$=$0 of 330.588\,GHz (e.g. Lovas 1992).
The difference in velocity is only 4.1\,km/s. Nevertheless, 
it is highly unlikely that the absorption line seen at
330.59\,GHz is caused by a Galactic $^{13}$CO(3--2) transition.
First of all, the depth of the absorption as well as the width of
the H$_2$O line is the same as those of the lines of redshifted
CO(2--1), $^{13}$CO(2--1), C$^{18}$O(2--1), HCO$^+$(2--1),
HCN(2--1), etc. (Combes \& Wiklind 1996) -- in itself a
strong indication that our new line is from redshifted water.
Secondly, a search through all our spectra, covering several
GHz, does not reveal any molecular transition at z$=$0, although
several relatively strong lines should be present (for instance:
SO($3_{4}-2_{3}$), SiO(3--2) $v=0$).
Thirdly, B0218+357 is situated at Galactic coordinates
$l=142.6^{\circ}$, $b=-23.5^{\circ}$. This means that unless
Galactic absorption occurs very locally, Galactic rotation would
displace the z$=$0 line of $^{13}$CO(3--2) to negative velocities.
If there is local gas, it is likely to be extended on scales of 1'
(the throw of our telescope beam) and the observing procedure with
a nutating subreflector would effectively cancel Galactic absorption.

Figure \ref{h2o_f2} displays our continuum measurements, together with
a compilation of previous results in the literature for lower frequencies.
Within the 1$\sigma$ error bars, the continuum spectrum can be fitted with
a power law of slope $-0.25$. From our previous detection of the 
C$^{18}$O(2--1) line with an 
optical depth of $\approx$ 3 (Combes \& Wiklind 1995), we deduced an optical
depth of 1500 for the $^{12}$CO(2--1) line. Since the H$_2$O abundance is likely
to be only 10 times lower than that of CO, while its dipole moment
$\mu$ = 1.8 debye is 18 times higher, we expect an H$_2$O optical depth about
30 times higher than CO for cold gas, since $\tau/N$ scales as $\mu^2$. This 
clearly prevents any estimation of the H$_2$O abundance; however the line
strength S is about ten times lower for the 183 Ghz line, so the upper limit
on the 183 GHz line provides a constraint on the excitation temperature.

The total column density of the H$_2$O molecule, 
observed in absorption between the levels $ l\longrightarrow u $ with an
optical depth $\tau$ at the center of the observed line of width $\Delta v$
at half--power is:
$$
N_{H_2O} = \alpha f(T_x) {{\nu^3 \tau \Delta v} \over {g_u s_I A_u} }\ ,
$$
where $\alpha$ is a constant (8$\pi$/c$^3$), $\nu$ is the frequency of the 
transition, $g_u$ the statistical weight of the upper level
(= 2 J$_u$+1), $A_u$ the Einstein coefficient of the transition, 
$T_x$ the excitation temperature, and 
$$
f(T_x) = {{Q(T_x) exp(E_l/kT_x)} \over { 1 - exp(-h\nu/kT_x)}}\ ,
$$
where $Q(T_x)$ is the partition function.
The factor $s_I$ is the nuclear spin statistical weight, equal to
\slantfrac{3}{4} for ortho states and \slantfrac{1}{4} for para states.
 
 For the sake of simplicity, we adopt the hypothesis of restricted 
thermodynamical equilibrium conditions, i.e. that the excitation
temperature is the same for all the H$_2$O lines.  Also, we
assume an ortho/para ratio of 3. Replacing in the 
above formula the molecular parameters from de Lucia, Helminger
\& Kirchoff (1974), 
and assuming the abundance of H$_2$O/CO $\approx$ 0.1, or
H$_2$O/H$_2 \approx$ 10$^{-5}$, which is found for the galactic ISM,
the optical depths of the two H$_2$O observed lines can be predicted
as displayed in Figure \ref{h2o_f3}. The 3$\sigma$ upper limit to
the 183 GHz line constraints then $T_x$ to be lower than 20 K.

\medskip

This result implies that the bulk of the H$_2$O molecules that we
detect in absorption are not coming from
hot dense cores, but are more widely spread and mixed with the
molecular cloud absorbing in CO. This is consistent with the high
covering factor observed, and with the fact that the absorption
technique selects preferentially cold gas (e.g. Combes \& Wiklind 1996;
Wiklind \& Combes 1997).
Also, the absorbing gas is situated in an intervening cloud which
happens to be on the line of sight of the remote quasar. It is thus
not necessarily an actively star forming region, as is the case for
emission line observations of distant galaxies.
It should be emphasized, however, that this result is based on the
assumption of H$_2$O galactic abundance, which is poorly known; another
solution could be a lower H$_2$O abundance, which will release the
constraint of low temperature. However, even with an abundance
of H$_2$O/H$_2$ = 10$^{-6}$ (or H$_2$O/CO=0.01), 
the excitation temperature should
be lower than 30K (see fig \ref{h2o_f3}). A higher H$_2$O abundance is 
not likely, unless we release the hypothesis of
a constant $T_x$ over the rotational ladder.

\acknowledgments
The present H$_2$O line detection at 331\,GHz could not have been done without 
the enthusiastic support from the IRAM staff at the Pico Veleta.
Bibliographic and photometric data have been retrieved from the NED data
base.

\clearpage

\begin{deluxetable}{lccl}
\small
\tablecaption{Observational parameters. \label{tbl-1}}
\tablewidth{0pt}
\tablehead{
\colhead{J$_{KaKc}$}                        &
\colhead{3$_{13}$-2$_{20}$}                 &
\colhead{1$_{10}$-1$_{01}$}
}
\startdata
$\nu_{lab}$\ GHz              & 183.310                      & 556.936  \nl
$\nu_{obs}$\ GHz              & 108.811                      & 330.592  \nl
\nl
Forward eff.\tablenotemark{a} &  0.92                        & 0.77     \nl
Beam eff.\tablenotemark{a}    &  0.74                        & 0.19     \nl
\nl
T$_A^*$                       & $<$ 2.5\,mK\tablenotemark{b} & 5.5\,mK  \nl
FWHM (km/s)                   & 15.                          & 15.      \nl
$\sigma$ (9km/s)              & 1.1\,mK                      & 1.3\,mK  \nl
\nl

\enddata

\tablenotetext{a}{Main--beam efficiency $\eta_{\rm mb} = B_{\rm eff}/F_{\rm eff}$}
\tablenotetext{b}{$^*$ 3$\sigma$ upper limit in 15km/s channels}
\end{deluxetable}

\clearpage

\figcaption[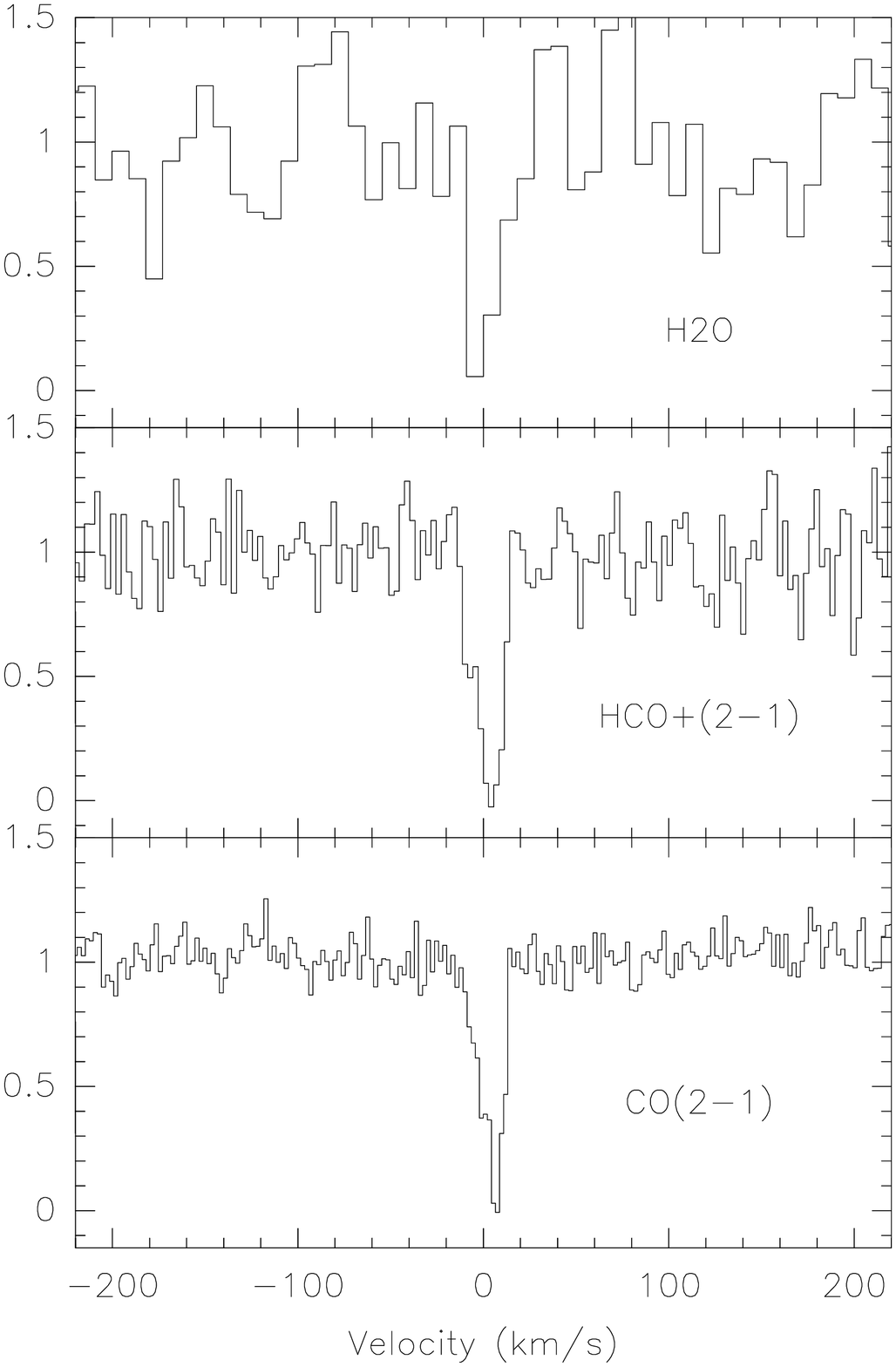]{Spectrum of ortho--water in its fundamental line
at 557\,GHz, redshifted at 331\,GHz, in absorption towards B0218+357
at z$=$0.68466.
The line has the same width as the previously detected HCO$^+$(2--1) and
CO(2--1) lines. The velocity resolution is 9.1, 2.8 and 2.2\,km/s from
top to bottom. Spectra have been normalised to the continuum level completely
absorbed (33\% of the total), i.e. 6, 37 and 27mK in $T_A^*$ scale respectively.
\label{h2o_f1}}
 
\figcaption[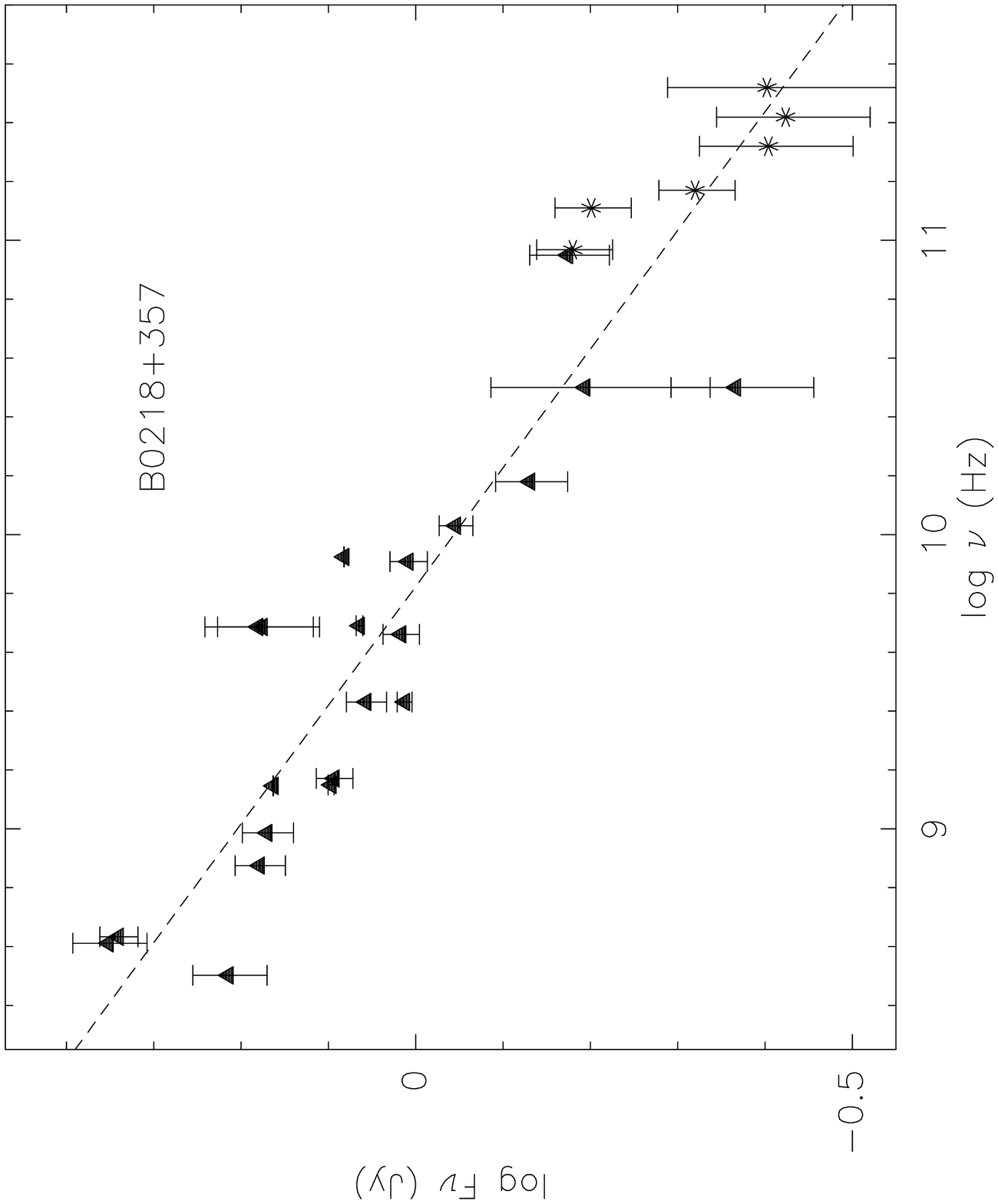]{Radio continuum spectrum of B0218+357. Measurements from
the literature (see NED for a compilation) are  plotted with filled triangles, 
and the stars correspond to the present work.
The dashed line is not a fit, but represents a power law of slope $-0.25$.
\label{h2o_f2}}
 
\figcaption[h20_f3.ps]{Logarithm of the central optical depth predicted for
the two observed H$_2$O lines, for an assumed abundance H$_2$O/H$_2$=10$^{-5}$
(full line), and 10$^{-6}$ (dashed line),
as  a function of excitation temperature. The horizontal line is the 3$\sigma$
upper limit for the 183\,GHz line, indicating that the excitation temperature
is less than 20\,K (or 30\,K, dashed line).
\label{h2o_f3}}
 
\end{document}